\documentclass[a4paper,11pt]{article}
\usepackage{jheppub}
\usepackage[T1]{fontenc}
\usepackage{graphicx,subcaption}
\usepackage{braket}
\usepackage{bm}
\usepackage{xcolor}

\usepackage{acro}
\DeclareAcronym{gr}{
    short=GR ,
    long=general relativity
}
\DeclareAcronym{ds}{
    short=dS ,
    long=de Sitter
}
\DeclareAcronym{ads}{
    short=AdS ,
    long=anti-de Sitter
}
\DeclareAcronym{qft}{
    short=QFT ,
    long={quantum field theory}
}
\DeclareAcronym{jt}{
    short=JT ,
    long={Jackiw-Teitelboim}
}

\DeclareAcronym{uv}{
    short=UV ,
    long=ultraviolet
}
\DeclareAcronym{ir}{
    short=IR ,
    long=infrared
}
\DeclareAcronym{wdw}{
    short=WDW ,
    long=Wheeler-DeWitt
}
\DeclareAcronym{adm}{
    short=ADM ,
    long={Arnowitt, Deser, and Misner}
}
\allowdisplaybreaks[1]

\title{\bf Functional Determinants for Constrained Path Integrals in Minisuperspace Jackiw-Teitelboim Gravity}

\author{\large Hiroki Matsui${}^{c,d}$}
\emailAdd{hiroki.matsui@yukawa.kyoto-u.ac.jp}

\affiliation{
${}^c$Department of Physics, College of Humanities and Sciences, Nihon University, Sakurajosui, Tokyo 156-8550, Japan \medskip\\
${}^d$Center for Gravitational Physics and Quantum Information, Yukawa Institute for Theoretical Physics, Kyoto University, Kitashirakawa Oiwakecho, Sakyo-ku, Kyoto 606-8502, Japan }

\abstract{ We present a detailed evaluation of constrained minisuperspace path integrals in Jackiw-Teitelboim (JT) gravity and in biaxial Bianchi IX quantum cosmology, employing the Gelfand-Yaglom theorem to compute the relevant functional determinants. In both settings, integrating out the dilaton or a minisuperspace variable produces a functional delta that enforces the classical constraint equation, thereby localizing the remaining path integral onto classical configurations. The associated Jacobian, equivalently, the functional determinant of the operator obtained by linearizing the constraint about the classical solution, fixes the semiclassical prefactor and the correct measure. We evaluate this determinant exactly via the Gelfand-Yaglom method and obtain the fully normalized fixed-lapse propagators. We further extend the JT analysis to a quadratic dilaton potential $U(\phi)=m^{2}\phi^{2}$ and comment on the corresponding saddle-point structure of the lapse integral. Finally, we apply the same approach to Bianchi IX quantum cosmology and derive the fixed-lapse
propagator, including its prefactor. Our results provide a systematic and broadly applicable prescription for treating constraint structures in gravitational path integrals using functional determinant techniques, with potential applications to a wider class of minisuperspace quantum cosmology and quantum gravity.
}
\begin{document}
\maketitle
\flushbottom

\section{Introduction}
\label{sec:introduction}

The quantization of gravity remains one of the most fundamental open problems in theoretical physics. In contrast to the other fundamental interactions, \ac{gr} resists straightforward quantization because of its diffeomorphism invariance, the absence of a fixed background structure, and its non-renormalizability when treated as an ordinary \ac{qft}~\cite{tHooft:1974toh}. These difficulties make it challenging to construct a conventional quantum theory of the spacetime metric itself. Nevertheless, the path integral formulation offers a conceptually appealing framework for quantum gravity, in which transition amplitudes are expressed as functional integrals over equivalence classes of spacetime geometries~\cite{Hawking:1978jz}.

The gravitational path integral in general form is written as
\begin{equation}\label{eq:G-amplitude1}
G(g_{f};g_{i})= \int\frac{\mathcal{D}g_{\mu\nu}}{\textrm{Vol(Diff)}} \exp \left(\frac{i S[g_{\mu\nu}]}{\hbar}\right)\,,
\end{equation}
where $S[g_{\mu\nu}]$ is the gravitational action and the integration is over all metric configurations interpolating between initial and final geometries ($g_i$ and $g_f$). The division by $\mathrm{Vol}(\mathrm{Diff})$ indicates the removal of diffeomorphism redundancy. For \ac{gr}, $S[g_{\mu\nu}]$ is taken to be the Einstein-Hilbert action supplemented by the appropriate boundary term~\cite{York:1972sj,Gibbons:1976ue}, which, in units $8\pi G = 1$, can be written as
\begin{equation}
S[g_{\mu\nu}]
= \frac{1}{2}\int_{\mathcal{M}} \mathrm{d}^4x \sqrt{-g}\,\bigl(R - 2 \Lambda\bigr)
  + \int_{\partial \mathcal{M}} \mathrm{d}^3y \sqrt{g^{(3)}}\,\mathcal{K}\,.
\end{equation}
In the canonical formulation, the amplitude \eqref{eq:G-amplitude1} may be viewed as formally generating solutions to the \ac{wdw} equation~\cite{DeWitt:1967ub}, which plays the role of the fundamental equation for the wave functional of the universe. However, as it stands, the functional integral \eqref{eq:G-amplitude1} over all possible  Lorentzian geometries is ill-defined and extremely difficult to evaluate in a mathematically controlled way, especially in four spacetime dimensions.

These issues motivate the study of simplified models that more rigorously define the gravitational path integral while retaining essential physical properties. Lower-dimensional gravity theories have played an important role in this regard, clarifying conceptual and non-perturbative aspects of quantum gravity. In particular, two-dimensional dilaton gravity captures key features of gravitational dynamics such as black hole thermodynamics, holographic AdS/CFT dualities, and non-perturbative effects while often remaining exactly solvable or useful to the approximation. Among these theories, \ac{jt} gravity~\cite{Jackiw:1984je,Teitelboim:1983ux} has attracted considerable attention. Most work has focused on \ac{jt} gravity in \ac{ads} spacetime~\cite{Almheiri:2014cka,Jensen:2016pah,Maldacena:2016upp,Engelsoy:2016xyb,
Stanford:2017thb,Saad:2019lba}, but there has also been significant progress in developing its \ac{ds} counterpart~\cite{Maldacena:2019cbz,Stanford:2019vob,Cotler:2019nbi}. The \ac{ds} version is particularly relevant for quantum cosmology, given the observed positive cosmological constant, and thus provides a natural testing ground for proposals about cosmological initial conditions and gravitational path integrals.

In what follows, we focus on \ac{jt} gravity as a toy model for exploring these questions. The \ac{jt} action couples the metric $g_{\mu\nu}$ to a dilaton field $\phi$ and is given by
\begin{equation}\label{eq:jt-action}
S_{\mathrm{JT}}= 
\frac{1}{2}\int_{\mathcal{M}}\mathrm{d}^2x\sqrt{-g}\,\phi(R-2\Lambda)
-\int_{\partial\mathcal{M}}\mathrm{d}y\sqrt{\gamma}\,\phi\,\mathcal{K},
\end{equation}
where $\Lambda$ is the cosmological constant, $\gamma$ is the induced metric on the boundary, and $\mathcal{K}$ is the corresponding extrinsic curvature. The boundary term ensures a well-posed variational principle for fixed boundary conditions. The dilaton field $\phi$ appears linearly and therefore acts as a Lagrange multiplier enforcing the constant-curvature constraint $R = 2\Lambda$. 
This constraint structure is precisely what makes \ac{jt} gravity an ideal model for studying the structure of the gravitational path integral. Under asymptotically AdS$_2$ boundary conditions, it is well established that the path integral reduces to the boundary Schwarzian theory~\cite{Maldacena:2016upp,Engelsoy:2016xyb,Stanford:2017thb,Saad:2019lba}.
Moreover, because the path integral over the dilaton $\phi$ can be carried out exactly, the remaining metric path integral is localized onto constant-curvature geometries, yielding a purely geometric representation of the \ac{jt} path integral. In particular, integrating out $\phi$ and then evaluating the resulting constrained path integral, we obtain the following formula,
\begin{equation}
\label{eq:jt-amplitude}
\begin{split}
G(g_{f},\phi_{f};g_{i},\phi_{i})
&= \int \frac{\mathcal{D}g_{\mu\nu}}{\textrm{Vol(Diff)}}\,
    \delta\!\left(R(g) - 2\Lambda\right)
    \exp\!\left(- \frac{i}{\hbar}\int_{\partial\mathcal{M}}\mathrm{d}y\sqrt{\gamma}\,\phi_{\rm bdy}\,\mathcal{K}\right) \\
&= \sum_{g_{\text{cl}}}
    \frac{\exp\!\left[-\tfrac{i}{\hbar}\left(\int_{\partial\mathcal{M}}\mathrm{d}y\sqrt{\gamma}\,\phi_{\rm bdy}\,\mathcal{K}\right)_{g_{\text{cl}}}\right]}
         {\left|\det\!\left(\delta (R-2\Lambda)/\delta g_{\mu\nu}\right)_{g_{\text{cl}}}\right|}\,,
\end{split}
\end{equation}
where $\delta[\cdots]$ denotes the functional delta, and $\phi_{\rm bdy}$ collectively denotes the dilaton values on $\partial\mathcal{M}$ (in particular the initial and final values $\phi_i$
and $\phi_f$ in the present amplitude).
In the last step, we have used the functional generalization of Eq.~\eqref{eq:functional-Dirac-delta} to evaluate the $\delta$-functional by expanding around the classical metric $g_{\text{cl}}$ satisfying $R(g_{\rm cl})=2\Lambda$ with the boundary conditions.
The delta functional in Eq.~\eqref{eq:jt-amplitude} must be carefully calculated and requires computing the determinant of the linearised constraint operator. 
Once the determinant has been evaluated, Eq.~\eqref{eq:jt-amplitude} yields the gravitational transition amplitude for an arbitrary Lorentzian geometry in \ac{jt} gravity. In practice, however, computing this determinant for fluctuations of the full metric $g_{\mu\nu}$ with an appropriate gauge fixing is technically challenging. A complete derivation and detailed analysis will be presented in a separate publication.

Instead, in this paper, we evaluate the constrained path integral in the minisuperspace approximation, which provides a practical approach for evaluating the path integral of quantum gravity. This minisuperspace approximation truncates the infinite degrees of freedom of the metric to a finite variable by assuming a high degree of symmetry (e.g., homogeneity and isotropy). In our previous collaborative work~\cite{Honda:2024hdr}, we evaluated the minisuperspace path integral for \ac{jt} gravity by performing the path integrals over the dilaton and the scale factor separately. This allowed us to evaluate the \ac{jt} path integral without resorting to the constrained path integral of Eq.~\eqref{eq:jt-amplitude}. In this paper, we first derive and evaluate the constrained path integral in minisuperspace and demonstrate that it yields the same results as those reported in~\cite{Honda:2024hdr}. Furthermore, to determine the prefactor in the path integral, we compute the functional determinant using the Gelfand-Yaglom theorem~\cite{Gelfand:1959nq}. This theorem provides a powerful method for computing determinants of one-dimensional differential operators by relating them directly to solutions of the corresponding classical equations of motion~\cite{Coleman:1985rnk,Kirsten:2004qv,Dunne:2007rt}. Thus, this work provides a full evaluation of the Lorentzian path integral in \ac{jt} gravity.
Then, we generalize our analysis to \ac{jt} gravity with dilaton potential $U(\phi)=m^2\phi^2$, which includes the pure \ac{jt} case as $m^2=0$. 
As a further application of our methods, we discuss the calculation for Bianchi IX quantum cosmology~\cite{Hawking:1984wn,Wright:1984wm,Amsterdamski:1985qu,Jensen:1990xc,DiazDorronsoro:2018wro,Feldbrugge:2018gin,Janssen:2019sex,Lehners:2024kus,LoFranco:2025myd}, which involves several minisuperspace variables and exhibits a more complex constraint structure. This demonstrates the broader applicability of functional determinant techniques in minisuperspace quantum cosmology and quantum gravity.

The paper is organized as follows.
In Section~\ref{sec:JT-minisuperspace}, we review the minisuperspace reduction of \ac{jt} gravity and rewrite the minisuperspace path integral as a constrained path integral by integrating out the dilaton.
In Section~\ref{sec:GY-theorem}, we present the general treatment of functional delta constraints, and compute the relevant functional determinant using the Gelfand-Yaglom theorem.
In Section~\ref{sec:lapse-integral}, we perform the lapse integral over $N>0$ to obtain the Lorentzian transition amplitude of \ac{jt} gravity, and review the results of~\cite{Honda:2024hdr} relevant for our discussion.
In Section~\ref{sec:general-JT}, we extend the analysis to general \ac{jt} gravity with a dilaton potential $U(\phi)=m^{2}\phi^{2}$.
In Section~\ref{sec:Bianchi-ix}, we apply our analysis to Bianchi IX quantum cosmology and compute the functional determinant. Section~\ref{sec:conclusion} concludes with a summary and future directions.

\section{Jackiw-Teitelboim Gravity}
\label{sec:JT-minisuperspace}

The \ac{jt} gravity provides the simplest example of two-dimensional dilaton gravity. We consider the minisuperspace reduction to one spatial dimension with metric $\mathrm{d}s^2=-N(t)^2\mathrm{d}t^2+a(t)^2\mathrm{d}x^2$, where $x$ has period $2\pi\ell$ with $\ell$ the spatial coordinate length and $N(t)$ is the lapse function. The minisuperspace \ac{jt} action takes the form~\cite{Honda:2024hdr}
\begin{equation}
  S_{\mathrm{JT}}=-\ell\int_{t_0}^{t_1}\!{\mathrm{d}}t 
  \left(\frac{\dot q\,\dot\varphi}{4N}+N\Lambda\right),
  \label{eq:jt-mini-action}
\end{equation}
defined on the time interval $[t_0,t_1]$ of length $T:=t_1-t_0>0$.
In what follows, we fix the gauge by choosing a constant positive lapse,
$N>0$ with $\dot N = 0$. We have introduced
\begin{equation}
q(t):=a(t)^2,\quad \varphi(t):=\phi(t)^2,
\end{equation}
and we transformed the lapse function $N \to N/a\phi$ (see Ref.~\cite{Honda:2024hdr} for the details).
We impose Dirichlet boundary conditions on both fields,
\begin{equation}
  q(t_0)=q_0,\quad q(t_1)=q_1,\quad 
  \varphi(t_0)=\varphi_0,\quad \varphi(t_1)=\varphi_1.
  \label{eq:Dirichlet-bc}
\end{equation}
These specify the initial and final geometric configurations, appropriate for computing transition amplitudes in quantum cosmology.
Once the boundary conditions are imposed, we can evaluate the gravitational path integral. Through the Batalin-Fradkin-Vilkovisky (BFV) formalism~\cite{Fradkin:1975cq,Batalin:1977pb}, the gravitational path integral~\eqref{eq:G-amplitude1} of \ac{jt} gravity reduces to a quantum mechanical-like path 
integral~\cite{Halliwell:1988wc,Halliwell:1988ik}
\begin{equation}\label{eq:G-amplitude2}
G(q_1,\varphi_1;q_0,\varphi_0)= \int \mathrm{d} N\left(t_1-t_0\right)\int \mathcal{D}q~\mathcal{D}\varphi~
\exp \left(\frac{i}{\hbar}S_{\mathrm{JT}}[N,q,\varphi]\right)~,
\end{equation}
which is the integral over the proper time $N\left(t_1-t_0\right)$ between the initial and final configurations. 
Much of the recent work in Lorentzian quantum cosmology~\cite{Feldbrugge:2017kzv,Lehners:2023yrj} has focused on giving a rigorous definition of the lapse integral $\int \mathrm{d}N$ using Picard-Lefschetz theory~\cite{Witten:2010cx}. This is crucial for resolving ambiguities between different cosmological proposals, such as the Hartle-Hawking no-boundary proposal~\cite{Hartle:1983ai} and Vilenkin's tunneling proposal~\cite{Vilenkin:1984wp}, as well as related early proposals~\cite{Linde:1983cm,Linde:1984ir,Rubakov:1984bh,Zeldovich:1984vk}.
However, to address this problem completely, we must combine resurgence theory with a systematic analysis of Lefschetz thimbles~\cite{Honda:2024aro}. Within this framework, the minisuperspace path integral can be evaluated consistently, and the ambiguity in the wave function of the universe can be resolved.

By performing the path integrals over the dilaton and the scale factor separately, we obtain the following propagator~\cite{Honda:2024hdr}, 
\begin{equation}\label{eq:jt-propagator}
 K(q_1, \varphi_1;q_0, \varphi_0) = \frac{\ell}{8\pi\hbar NT}\exp\!\left[\frac{i}{\hbar}\left(-\ell\,N\Lambda T
    - \ell\,\frac{q_1-q_0}{4NT}\,(\varphi_1-\varphi_0)\right)\right]\,, 
\end{equation}
and the minisuperspace transition amplitude for \ac{jt} gravity is written as 
\begin{equation}
\label{eq:jt-transition-amplitude-0}
G(q_1,\varphi_1;q_0,\varphi_0)
=\int_{0}^{\infty}\!{\mathrm{d}}NK(q_1,\varphi_1;q_0,\varphi_0)\,.
\end{equation}
When we have integrated the propagator for $N$ over all
positive values of the lapse function, we can get the Lorentzian transition amplitude of \ac{jt} gravity. The detailed discussion is given by Ref.~\cite{Fanaras:2021awm,Honda:2024hdr}, and we provide a brief review in Section~\ref{sec:lapse-integral}.

On the other hand, we can transform the gravitational path integral~\eqref{eq:G-amplitude2} into a constrained path integral and then explicitly perform the integration. 
Performing an integration by parts on 
the kinetic term in Eq.~\eqref{eq:jt-mini-action} yields
\begin{equation} S_{\mathrm{JT}} =\ell\int_{t_0}^{t_1}\!{\mathrm{d}}t\,\varphi\,\frac{\ddot q}{4N} -\ell\int_{t_0}^{t_1}\!{\mathrm{d}}t\,N\Lambda -\ell\left[\frac{\dot q}{4N}\,\varphi\right]_{t_0}^{t_1}. \label{eq:action-IBP} 
\end{equation}
The bulk term now makes it explicit that the dilaton $\varphi$ plays the role of a Lagrange multiplier enforcing the constraint $\ddot q = 0$. Since $\varphi$ appears only linearly in the exponent, the $\varphi$ path integral is of Fourier type rather than Gaussian and produces a functional delta enforcing the constraint,
\begin{align}
\begin{split}
  \int \mathcal{D}q\,\mathcal{D}\varphi\,
  e^{\frac{i}{\hbar}S_{\mathrm{JT}}[q,\varphi,N]}
  &=\int\!\mathcal{D}q\,\delta\!\left(\frac{\ddot q}{4N}\right)
    \exp\!\left[\frac{i}{\hbar}\left(
     -\ell\!\int_{t_0}^{t_1}\!{\mathrm{d}}t\,N\Lambda
     -\ell\left[\frac{\dot q}{4N}\,\varphi\right]_{t_0}^{t_1}\right)\right] \\
  &=\int\!\mathcal{D}q\,\delta\!\left(\frac{\ddot q}{4N}\right)
    \exp\!\left[\frac{i}{\hbar}S_{\textrm{on-shell}}[q,N]\right] ,
\end{split}
  \label{eq:jt-path-integral}
\end{align}
where $S_{\textrm{on-shell}}[q,N]$ denotes the on-shell action obtained after integrating out $\varphi$, and includes the cosmological term together with the boundary contributions. This path-integral representation should reproduce the propagator in Eq.~\eqref{eq:jt-propagator}. It also makes manifest a key structural feature of \ac{jt} gravity: the dilaton does not propagate as an independent dynamical degree of freedom, but instead acts as a Lagrange multiplier imposing a constraint on the metric. Consequently, the gravitational path integral~\eqref{eq:G-amplitude2} of \ac{jt} gravity can be transformed into a constrained path integral
with a functional delta. 
\footnote{Although we aim to verify and compute the constrained path integral of \ac{jt} gravity~\eqref{eq:jt-amplitude}, to make the method presented in this paper more transparent, we make use of the gravitational path integral~\eqref{eq:G-amplitude2} with the \ac{jt} action~\eqref{eq:jt-mini-action}  to illustrate the analysis of such constrained path integrals. In fact, when we take metric $\mathrm{d}s^2=-(N^2/q(t)\varphi(t)) \mathrm{d}t^2+q(t)\mathrm{d}x^2$ and insert it into the action~\eqref{eq:jt-action} and Eq.~\eqref{eq:jt-amplitude}, we can derive the constrained path integral~\eqref{eq:jt-path-integral}.}

A central technical difficulty in evaluating such constrained path integrals is the proper treatment of the functional delta and its associated Jacobian. Performing a functional delta function gives rise to functional determinants of differential operators, which must be carefully calculated. The Gelfand-Yaglom theorem~\cite{Gelfand:1959nq} provides an elegant way to compute these determinants: for a differential operator defined on a finite interval with specified boundary conditions, the theorem expresses its functional determinant in terms of the boundary value of the solution to an associated initial-value problem. This method has been widely used in \ac{qft}: detailed reviews can be found in Refs.~\cite{Coleman:1985rnk,Dunne:2007rt}. In the following, we use the Gelfand-Yaglom theorem to analyze the constrained path integral.
\footnote{Another study~\cite{Ailiga:2024wdx} employs the Gelfand-Yaglom method to compute the functional determinants in Lorentzian quantum cosmology.}

\section{Functional Determinants and Gelfand-Yaglom Theorem}
\label{sec:GY-theorem}

In this section, we explain how to evaluate functional determinants using the Gelfand-Yaglom theorem. First, let us recall the corresponding structure for an ordinary function. For a function $F(x)$, the Dirac delta function satisfies
\begin{equation}
\int \mathrm{d}x\,\delta\left(F(x)\right)\psi(x)
= \sum_{x_i:\,F(x_i)=0} 
    \frac{\psi(x_i)}{\big|F'(x_i)\big|},
\end{equation}
where the sum extends over all zeros $x_i$ of $F(x)$, and the Jacobian factor $\bigl|F'(x_i)\bigr|$ ensures the correct transformation of the measure under the change of variables.
The functional generalization of this formula reads
\begin{align}\label{eq:functional-Dirac-delta}
 \int\!\mathcal{D}q\,\delta\left(\mathcal{G}[q]\right)\Psi[q] = \sum_{q_{\mathrm{cl}}:\,\mathcal{G}[q_{\mathrm{cl}}]=0} 
  \frac{\Psi[q_{\mathrm{cl}}]}
       {\left|\det\!\left(\delta \mathcal{G}/\delta q\right)_{q_{\mathrm{cl}}}\right|},
\end{align}
where the sum runs over all configurations $q_{\mathrm{cl}}$ satisfying the functional constraint $\mathcal{G}[q_{\mathrm{cl}}]=0$. 
The functional derivative $\delta\mathcal{G}/\delta q$ at $q_{\mathrm{cl}}$ is defined via the first-order functional expansion
\begin{equation}
\label{eq:G-expansion}
\mathcal{G}[q_{\mathrm{cl}}+\delta q]
=\mathcal{G}[q_{\mathrm{cl}}]
 +\left(\frac{\delta \mathcal{G}}{\delta q}\right)_{q_{\mathrm{cl}}}\delta q
 +\mathcal{O}(\delta q^2)\,,
\end{equation}
so that $(\delta \mathcal{G}/\delta q)_{q_{\mathrm{cl}}}$ is understood as a linear operator acting on the fluctuation $\delta q$.

For the minisuperspace reduction of \ac{jt} gravity, the constraint takes the simple form
\begin{equation}
\mathcal{G}[q] = \frac{\ddot q}{4N}\,.
\end{equation}
Linearizing around a classical configuration $q_{\mathrm{cl}}$ satisfying $\ddot{q}_{\mathrm{cl}}=0$, we obtain
\begin{equation}
\delta \mathcal{G}
= \left(\frac{1}{4N}\,\partial_t^2\right)\delta q\,.
\end{equation}

The operator appearing in the determinant is thus $\mathcal{O}=(1/4N)\partial_t^2$, acting on functions satisfying Dirichlet boundary conditions $\delta q(t_0)=\delta q(t_1)=0$. 
With these boundary conditions on the interval $[t_0,t_1]$, the constraint $\ddot{q}=0$ admits a unique classical solution,
\begin{equation}
q_{\mathrm{cl}}(t)
 = q_0 + \frac{q_1 - q_0}{T}\,(t - t_0)\,.
\label{eq:classical-path}
\end{equation}
Using the functional delta representation \eqref{eq:functional-Dirac-delta}, the constrained path integral reduces to
\begin{equation}
\int\!\mathcal{D}q\,\delta\left(\frac{\ddot q}{4N}\right)\,
e^{\frac{i}{\hbar}S_{\mathrm{JT}}[q,N]}= \frac{e^{\frac{i}{\hbar}S_\textrm{on-shell}[N]}}{\left|\det\!\left(\frac{1}{4N}\partial_t^2\right)_{D}\right|}\,,
\label{eq:delta-det-again}
\end{equation}
where the subscript $D$ indicates that the determinant is taken over functions satisfying Dirichlet boundary conditions, and $S_{\mathrm{on\text{-}shell}}[N]$ is the \ac{jt} action evaluated on the classical path \eqref{eq:classical-path},
\begin{equation}\label{eq:on-shell-action}
S_{\textrm{on-shell}}[N]=-\ell N\Lambda T-\ell\frac{q_1-q_0}{4NT}(\varphi_1-\varphi_0)\,.
\end{equation}

\subsection{General Constraints}

For \ac{jt} gravity, the constraint functional $\mathcal{G}[q]$ takes a simple form; however, for later convenience, we first treat a more general constraint.

We consider the following constraint
\begin{equation}
\mathcal{G}[q]=A(q)\ddot q\,.
\label{eq:G_general1}
\end{equation}
Expanding $q=q_{\rm cl}+\delta q$ around a classical solution $q_{\rm cl}$ that satisfies $\mathcal{G}[q_{\rm cl}]=0$ with Dirichlet boundary conditions on $[t_0,t_1]$, the first variation reads
\begin{equation}
\delta \mathcal{G}
=\left(A(q_{\rm cl})\partial_t^2+A'(q_{\rm cl})\ddot q_{\rm cl}\right)\,\delta q
=\left(A(q_{\rm cl})\partial_t^2\right)\,\delta q\,,
\label{eq:deltaG_general1}
\end{equation}
where we used $\ddot q_{\rm cl}=0$.
Thus, for the constraint~\eqref{eq:G_general1}, the operator in the determinant is written as,
\begin{equation}
\mathcal{O} \equiv A(q_{\rm cl})\partial_t^2\,.
\label{eq:O_general1}
\end{equation}

Next, we consider the more general constraint
\begin{equation}
\mathcal{G}[q]=A(q)\ddot q+B(q)\,.
\label{eq:G_general2}
\end{equation}
Expanding $q=q_{\rm cl}+\delta q$ around a classical solution $q_{\rm cl}$ that satisfies $\mathcal{G}[q_{\rm cl}]=0$ with Dirichlet boundary conditions on $[t_0,t_1]$, the first variation reads
\begin{equation}
\delta \mathcal{G}
=\left(A(q_{\rm cl})\partial_t^2+A'(q_{\rm cl})\ddot q_{\rm cl}+B'(q_{\rm cl})\right)\,\delta q\,.
\label{eq:deltaG_general2}
\end{equation}
Thus, for the constraint~\eqref{eq:G_general2}, the operator in the determinant is 
\begin{equation}
\mathcal{O} \equiv A(q_{\rm cl})\partial_t^2+A'(q_{\rm cl})\ddot q_{\rm cl}+B'(q_{\rm cl})\,.
\label{eq:O_general2}
\end{equation}
Although it involves somewhat intricate calculations, we can define operators as above to compute determinants for more general constraints.

\subsection{Gelfand-Yaglom theorem}

The Gelfand-Yaglom theorem~\cite{Gelfand:1959nq} states that for a one-dimensional
second-order differential operator of the form
\begin{equation}
\mathcal{O} \equiv -\partial_t^2 + V(t)
\end{equation}
defined on the interval $[t_0,t_1]$ with Dirichlet boundary conditions, the functional determinant can be expressed in terms of the solution of an associated
initial-value problem. More precisely, if we get $y(t)$ by solving the equation
\begin{equation}
-y''(t) + V(t)\,y(t) = 0,
 \quad
 y(t_0) = 0,\quad y'(t_0) = 1\,,
\end{equation}
we obtain
\begin{equation}
\det\bigl(-\partial_t^2 + V(t)\bigr)_{D}
\propto y(t_1)\,.
\end{equation}

In the \ac{jt} gravity case, the relevant operator is
\begin{equation}
\mathcal{O} = \frac{1}{4N}\,\partial_t^2\,,
\end{equation}
which is of the Gelfand-Yaglom form with $V(t)=0$ and 
acts on functions defined on $[t_0,t_1]$ with Dirichlet boundary
conditions at the endpoints.
The corresponding Gelfand-Yaglom problem is
\begin{equation}
 y''(t) = 0,
 \quad
 y(t_0) = 0,\quad y'(t_0) = 1\,,
\end{equation}
whose solution is simply
\begin{equation}
y(t) = t - t_0\,.
\end{equation}
Therefore, we get
\begin{equation}
\det\bigl(\partial_{t}^2\bigr)_{D}
\propto y(t_1)
= t_1 - t_0
= T\,.
\end{equation}
This implies
\footnote{
To consider the Gelfand-Yaglom problem in the canonical form, it is convenient to absorb the overall factor in the operator by a rescaling of the time coordinate. 
For the operator $\mathcal{O}=\frac{1}{4N}\partial_t^2$, we introduce
\begin{equation}
\bar t := 2\sqrt{N}\,t,
\quad
\bar t_{0,1}:=2\sqrt{N}\,t_{0,1},
\end{equation}
so that
\begin{equation}
\mathcal{O}=\partial_{\bar t}^2
\end{equation}
acting on functions defined on
$[\bar t_0,\bar t_1]$ with Dirichlet boundary conditions at the endpoints.
Hence, we get
\begin{equation}
\det\!\left(\partial_{\bar t}^2\right)_{D}\propto y(\bar t_1)
=\bar t_1-\bar t_0
=2\sqrt{N}\,T.
\end{equation}
In the final propagator, any overall normalization associated with constant rescalings of the operator is fixed unambiguously by the normalization
condition, e.g., the composition law, so that the correctly normalized result is independent of the particular normalization procedure. }
\begin{equation}
\left|\det\!\left(\frac{1}{4N}\,\partial_t^2\right)_{D}\right|
\propto T\,.
\label{eq:det-final}
\end{equation}

Substituting Eq.~\eqref{eq:det-final} into Eq.~\eqref{eq:delta-det-again}, 
we obtain the following expression for the propagator
\begin{equation}
 K(q_1, \varphi_1; q_0, \varphi_0)
 = \frac{\mathcal{C}}{T}\,
   \exp\!\left[\frac{i}{\hbar}\left(
      - \ell\,N\Lambda T
      - \ell\,\frac{q_1 - q_0}{4 N T}\,(\varphi_1 - \varphi_0)
   \right)\right],
\label{eq:propagator-unnormalized}
\end{equation}
where $\mathcal{C}$ is an overall normalization constant,
whose value will be fixed by an appropriate normalization condition on the
transition amplitude.

The normalization constant $\mathcal{C}$ is fixed by the composition law
\begin{align} 
\int \mathrm{d}q\, \mathrm{d}\varphi\; K(q_1, \varphi_1;q, \varphi)\;
K(q, \varphi;q_0, \varphi_0)
= K(q_1, \varphi_1;q_0, \varphi_0) \,. 
\end{align}
By using the composition law, we obtain 
\begin{equation}
\mathcal{C}^2\cdot \frac{8\pi\hbar N}{l(t_1-t_0)} = 
\frac{\mathcal{C}}{(t_1-t_0)}\ \Longrightarrow \ \mathcal{C}
=\frac{l}{8\pi\hbar N}\,.
\end{equation}
The propagator can be written as 
\begin{equation}\label{eq:jt-propagator-final}
 K(q_1, \varphi_1;q_0, \varphi_0) = \frac{l}{8\pi\hbar N(t_1-t_0)}\exp\!\left[\frac{i}{\hbar}\left(-\,\ell\,N\Lambda(t_1-t_0)
    - \,\ell\,\frac{q_1-q_0}{4N(t_1-t_0)}\,(\varphi_1-\varphi_0)\right)\right]\,,
\end{equation}
which coincides with the result \eqref{eq:jt-propagator} obtained in Ref.~\cite{Honda:2024hdr}. Thus, we have confirmed that our evaluation of the constrained path integral using the Gelfand-Yaglom theorem to compute the functional determinant is justified.

\section{Lorentzian Transition Amplitude and Saddle-Point Structure}
\label{sec:lapse-integral}

Having obtained the propagator at fixed lapse $N$, we construct the full Lorentzian transition amplitude by integrating over the positive lapse function, $N>0$. This lapse integral exhibits a nontrivial saddle-point structure, whose contributing saddles and associated steepest-descent contours encode characteristic features of Lorentzian quantum cosmology~\cite{Feldbrugge:2017kzv,Honda:2024aro}. A detailed analysis in the context of \ac{jt} gravity is presented in Ref.~\cite{Fanaras:2021awm,Honda:2024hdr}; here we summarize the results relevant for our discussion.

We again rewrite the Lorentzian transition amplitude, which is written as
\begin{equation}
\label{eq:jt-transition-amplitude}
G(q_1,\varphi_1;q_0,\varphi_0)
=\int_{0}^{\infty}\!{\mathrm{d}}NK(q_1,\varphi_1;q_0,\varphi_0)\,.
\end{equation}
The restriction to $N>0$ enforces causality, ensuring that a positive lapse ensures forward time evolution. We introduce the new parameters
\begin{equation}
\label{eq:beta-gamma-def}
\beta:=\ell\,\Lambda\,T, 
\quad
\gamma:=\frac{\ell}{4T}\,(q_1-q_0)(\varphi_1-\varphi_0),
\end{equation}
which depends on the cosmological constant, time interval, and boundary conditions. 
By using the change of variables $N=\sqrt{\gamma/\beta}\,e^{x}$ with principal branch and $x\in\mathbb{R}$ for $\beta\gamma>0$, the amplitude becomes
\begin{align}
G(q_1,\varphi_1;q_0,\varphi_0)
&=\frac{\ell}{8\pi\hbar\,T}\int_{0}^{\infty}\frac{{\mathrm{d}}N}{N}
\exp\!\left[-\frac{i}{\hbar}\left(\beta N+\frac{\gamma}{N}\right)\right] \nonumber\\[2pt]
&=\frac{\ell}{8\pi\hbar\,T}\int_{-\infty}^{\infty}\!{\mathrm{d}}x\;
\exp\!\left[-\frac{2i}{\hbar}\sqrt{\beta\gamma}\,\cosh x\right] \nonumber\\[2pt]
&=\frac{\ell}{4\pi\hbar\,T}\,K_{0}\!\left(\frac{2i}{\hbar}\sqrt{\beta\gamma}\right)
=-\,\frac{i\,\ell}{8\,T}\,H^{(2)}_{0}\!\left(\frac{2}{\hbar}\sqrt{\beta\gamma}\right),
\label{eq:JT-amp-K0}
\end{align}
where we used the integral representation $K_0(z)=\tfrac12\int_{-\infty}^{\infty}{\mathrm{d}}s\,e^{-z\cosh s}$ for the modified Bessel function and the relation $K_0(iz)=-(\pi i/2)\,H_0^{(2)}(z)$ to the Hankel function.

The asymptotic behavior of Eq.~\eqref{eq:JT-amp-K0} in the semiclassical limit $\hbar\to0$ reveals two distinct physical regimes:

\paragraph{Classical propagation ($\boldsymbol{\beta\gamma>0}$).}
When $\Lambda>0$ and $(q_1-q_0)(\varphi_1-\varphi_0)>0$, the amplitude is oscillatory:
\begin{equation}
G(q_1,\varphi_1;q_0,\varphi_0)\sim 
\frac{\ell}{4\pi\hbar\,T}\sqrt{\frac{\pi\hbar}{4\sqrt{\beta\gamma}}}\;
\exp\!\left[-\frac{2i}{\hbar}\sqrt{\beta\gamma}-\frac{\pi i}{4}\right]\,.
\label{eq:WKB-osc}
\end{equation}
The phase oscillates, and this regime corresponds to classical Lorentzian evolution.

\paragraph{Quantum tunneling ($\boldsymbol{\beta\gamma<0}$).}
When $\Lambda>0$ and $(q_1-q_0)(\varphi_1-\varphi_0)<0$, we obtain exponential suppression:
\begin{equation}
G(q_1,\varphi_1;q_0,\varphi_0)=\frac{\ell}{4\pi\hbar\,T}\,K_{0}\!\left(\frac{2}{\hbar}\sqrt{|\beta\gamma|}\right)
\sim 
\frac{\ell}{4\pi\hbar\,T}\sqrt{\frac{\pi\hbar}{4\sqrt{|\beta\gamma|}}}\;
e^{-\frac{2}{\hbar}\sqrt{|\beta\gamma|}}.
\label{eq:WKB-tun}
\end{equation}
This exponential suppression indicates quantum tunneling through a classically forbidden region, analogous to its quantum mechanics.

To understand the saddle-point structure more clearly, we use Lefschetz thimble techniques~\cite{Witten:2010zr,Cristoforetti:2012su}. Changing variables to $N=e^{x}$ with $x\in\mathbb{C}$, the exponent becomes
\begin{equation}
F(x):=\frac{i}{\hbar}\left[-\beta e^{x}-\frac{\gamma}{e^{x}}\right].
\end{equation}
Thus, the saddle points are given by 
\begin{equation}
x^{(m)}_{s\pm}=\tfrac12\log\!\left(\frac{\gamma}{\beta}\right)\pm i\pi m,
\quad m\in\mathbb{Z},
\quad\Longrightarrow\quad
N^{(m)}_{s\pm}=\pm\sqrt{\frac{\gamma}{\beta}}.
\label{eq:saddles-JT}
\end{equation}

\paragraph{Case $\boldsymbol{\beta\gamma>0}$.}
All relevant saddles have $\mathrm{Re}\,x_s\in\mathbb{R}$. The steepest-descent path through $x^{(0)}_{s+}=\tfrac12\log(\gamma/\beta)$ is equivalent to the original real-$x$ contour. The dominant contribution comes from
\begin{equation}
N^{(0)}_{s+}=+\sqrt{\frac{\gamma}{\beta}}>0,
\end{equation}
a real positive lapse, confirming that the dominant histories are classical Lorentzian evolutions.

\paragraph{Case $\boldsymbol{\beta\gamma<0}$.}
The naive real contour crosses Stokes lines. To resolve this, we perform a small shift $\hbar\to|\hbar|e^{i\theta}$ with small $0<|\theta|\ll 1$ and this shift selects a unique contributing saddle,
\begin{equation}
x^{(0)}_{s+}=\frac12\log\!\left|\frac{\gamma}{\beta}\right|-\frac{i\pi}{2}
\quad \Longleftrightarrow \quad
N^{(0)}_{s+}=i\sqrt{\frac{|\gamma|}{\beta}},
\label{eq:euclid-saddle}
\end{equation}
which lies along the imaginary axis (${\rm Im}N<0$). This represents a Euclidean saddle point in the complex lapse plane so that the path integral localizes onto a smooth complex geometry realizing the Lorentzian continuation of Euclidean evolution. This precisely captures the picture of Hartle-Hawking no-boundary proposal~\cite{Hartle:1983ai}.

\section{General JT Gravity with Dilaton Potential}
\label{sec:general-JT}

We now extend our analysis to the general \ac{jt} action including a dilaton potential, 
\begin{equation}
\label{eq:jt-action-general}
S_{\text{JT}}
= \frac{1}{2}\int_{\mathcal{M}}\mathrm{d}^2x\,\sqrt{-g}\;\phi\bigl[R - 2U(\phi)\bigr]
-\int_{\partial \mathcal{M}}\mathrm{d}y\,\sqrt{\gamma}\,\phi\,\mathcal{K}\,,
\end{equation}
where $U(\phi)$ is a dilaton potential. 
Specializing to the minisuperspace truncation introduced in the previous section, the action reduces to
\begin{equation}
\label{eq:jt-action-general-mini}
S_{\rm JT}
= -\ell\int_{t_0}^{t_1}\mathrm{d}t\,
\left(\frac{\dot q\,\dot\varphi}{4N} + N\,U(\varphi^{1/2})\right)\,.
\end{equation}

As a concrete example with physical motivation, we take a quadratic potential $U(\phi)=m^2\phi^2$, giving $U(\varphi^{1/2})=m^2\varphi$ and get
\begin{equation}
\label{eq:jt-general-action-mini}
S_{\rm JT}
= -\ell \int_{t_0}^{t_1}\mathrm{d}t\,
\left(\frac{\dot q\,\dot\varphi}{4N} + N m^2 \varphi\right).
\end{equation}

Integration by parts yields
\begin{equation}
\label{eq:action-IBP-general}
S_{\rm JT}
= \ell\int_{t_0}^{t_1}\mathrm{d}t\,
\varphi\left(\frac{\ddot q}{4N}-N m^2\right)
-\ell\left[\frac{\dot q}{4N}\,\varphi\right]_{t_0}^{t_1}.
\end{equation}
Performing the path integral over $\varphi$ therefore enforces the constraint
\begin{equation}
\label{eq:G-constraint-jt}
\mathcal{G}[q]\equiv \frac{\ddot q}{4N} - N m^2 = 0 \,,
\end{equation}
where the classical trajectory satisfies $\ddot q = 4N^2 m^2$.
With Dirichlet boundary conditions $q(t_0)=q_0$ and $q(t_1)=q_1$, 
the constraint
\eqref{eq:G-constraint-jt} admits a unique solution,
\begin{equation}
q_{\rm cl}(t)=q_0+v_0\left(t-t_0\right)
+2N^2m^2\left(t-t_0\right)^2,
\end{equation}
with the initial velocity fixed by the endpoints as
\begin{equation}
v_0=\frac{q_1-q_0}{T}-2N^2m^2T.
\end{equation}
This represents uniformly accelerated motion, contrasting with the linear evolution in the $m^2=0$ case.
The on-shell action is given by
\begin{align}
\label{eq:S-onshell-general}
S_{\textrm{on-shell}}[N]
&= -\frac{\ell}{4N}\left(\varphi_1\,\dot q_{\rm cl}(t_1)-\varphi_0\,\dot q_{\rm cl}(t_0)\right)\nonumber\\
&= -\frac{\ell}{4NT}(q_1-q_0)(\varphi_1-\varphi_0)
-\frac{\ell N T m^2}{2}\,(\varphi_1+\varphi_0),
\end{align}

The functional derivative of constraint has the same form as before: since the constraint is $\mathcal{G}[q]=A(q)\ddot{q}+B(q)$ with $A(q)=1/(4N)$ and $B(q)=-Nm^2$ both independent of $q$, we have $A'(q_{\rm cl})=B'(q_{\rm cl})=0$. 
Therefore, by using Eq.~\eqref{eq:O_general2} we obtain 
\begin{equation}
\label{eq:det-general}
\left|\det\!\left(\frac{\delta \mathcal{G}}{\delta q}\right)_{q_{\rm cl}}\right|
= \left|\det\!\left(\frac{1}{4N}\,\partial_t^2\right)_{D}\right|
\propto T\,,
\end{equation}
which coincides with Eq.~\eqref{eq:det-final}.
Implementing the normalization via the composition law, as in the pure \ac{jt} case, we obtain the propagator
\begin{equation}\label{eq:jt-propagator-general}
K(q_1, \varphi_1;q_0, \varphi_0) =
\frac{\ell}{8\pi\hbar NT}\exp\!\left[\frac{i}{\hbar}\left(-\frac{\ell m^2 NT}{2}\left(\varphi_1+\varphi_0\right)
    -\frac{\ell(q_1-q_0)}{4NT}\,(\varphi_1-\varphi_0)\right)\right]
\end{equation}
Setting $m^2=0$ correctly reproduces the pure \ac{jt} propagator \eqref{eq:jt-propagator-final} with $\Lambda=0$.
Moreover, the lapse dependence remains of the standard form, 
so the associated thimble structure is qualitatively the same as in pure \ac{jt} gravity.

\section{Bianchi IX Quantum Cosmology}
\label{sec:Bianchi-ix}

To demonstrate the broader applicability of our methods, we consider the application to Bianchi IX quantum cosmology. This model contains several independent degrees of freedom and provides a standard approach for studying anisotropic dynamics in quantum cosmology~\cite{Hawking:1984wn,Wright:1984wm,Amsterdamski:1985qu,Jensen:1990xc,DiazDorronsoro:2018wro,Feldbrugge:2018gin,Janssen:2019sex,Lehners:2024kus,LoFranco:2025myd}. The Bianchi IX model describes a homogeneous but anisotropic spacetime with spatial topology $S^{3}$. In the biaxial set-up, the metric can be parametrized by two time-dependent scale factors, which we denote by $p(t)$ and $q(t)$: roughly speaking, $p$ controls the overall volume while $q/p$ measures the anisotropy. The corresponding minisuperspace action reads~\cite{DiazDorronsoro:2018wro}
\begin{equation} 
S_{\mathrm{BIX}}=2\pi^2\int_{t_0}^{t_1}\mathrm{d}t\left[-\frac{1}{4 N}\left(\frac{q \dot{p}^2}{p}+2 \dot{p} \dot{q}\right)+N\left(4-\frac{q}{p}-p \Lambda\right)\right], \label{eq:BIX-action} 
\end{equation}
defined on a finite interval $t\in[t_0,t_1]$ with $T:=t_1-t_0>0$. We impose Dirichlet boundary conditions on both minisuperspace variables,
\begin{equation}
p(t_0)=p_0,\quad p(t_1)=p_1,\quad
q(t_0)=q_0,\quad q(t_1)=q_1,
\end{equation}
which specify the initial and final three-geometries.

\subsection{Bianchi IX Path Integral and Constraint Structure}

After integrating by parts, the action~\eqref{eq:BIX-action} can be written as
\begin{align}
S_{\mathrm{BIX}}
&=2\pi^2\int_{t_0}^{t_1}\mathrm{d}t\,q(t)\left[-\frac{1}{4N}\left(\frac{\dot p^{\,2}}{p}-2\ddot p\right)-\frac{N}{p}\right] \nonumber\\
&\quad+2\pi^2\int_{t_0}^{t_1}\mathrm{d}t\,N\,(4-\Lambda p)
+\frac{\pi^2}{N}\left[-q\,\dot p\right]_{t_0}^{t_1}.
\label{eq:BIX-IBP}
\end{align}
The variable $q(t)$ appears linearly in the action and therefore plays the role of a Lagrange multiplier: path integration over $q(t)$ produces a functional delta imposing the constraint on $p(t)$. Thus, we obtain 
\begin{align}
\begin{split}
K(q_1,p_1;q_0,p_0)
&=\int_{p(t_0)=p_0}^{p(t_1)=p_1}\!\!\mathcal{D}p\;
\delta\!\left(\mathcal{G}[p]\right)\\
&\quad\times
\exp\!\left[\frac{i}{\hbar}\left(
2\pi^2\!\int_{t_0}^{t_1}\!\mathrm{d}t\,N\,(4-\Lambda p)
+\frac{\pi^2}{N}\,[-q\dot p]_{t_0}^{t_1}\right)\right],
\end{split}
\label{eq:K-BIX}
\end{align}
where the constraint functional is
\begin{equation}
\mathcal{G}[p]\equiv -\frac{1}{4N}\left(\frac{\dot p^{\,2}}{p}-2\ddot p\right)-\frac{N}{p}=0.
\label{eq:Bianchi-constraint}
\end{equation}

Equation~\eqref{eq:Bianchi-constraint} is equivalent to the second-order ODE
\begin{equation}
2\ddot p=\frac{\dot p^{2}}{p}+\frac{4N^2}{p}.
\label{eq:ode-p}
\end{equation}

We introduce $u:=\dot p/p$ so that $\dot p=up$ and $\ddot p=p\dot u+u^2p$.
Thus, Eq.~\eqref{eq:ode-p} becomes
\begin{equation}
2\dot u+u^2=\frac{4N^2}{p^2}.
\end{equation}
Changing variables from $t$ to $p$ using $\dot u=(\mathrm{d}u/\mathrm{d}p)\dot p=(\mathrm{d}u/\mathrm{d}p)\,u p$ and setting $w:=u^2$, 
we find a first-order ODE for $w(p)$,
\begin{equation}
\frac{\mathrm{d}}{\mathrm{d}p}w+\frac{1}{p}w-\frac{4N^2}{p^3}=0
\quad\Longrightarrow \quad
w(p)=\frac{\alpha}{p}-\frac{4N^2}{p^2},
\label{eq:first-int}
\end{equation}
with integration constant $\alpha$. Equivalently,
\begin{equation}
\dot p=\pm\sqrt{\alpha p-4N^2}\,.
\label{eq:first-order}
\end{equation}

For $\alpha\neq 0$, the general solution is
\begin{equation}
p_{\rm cl}(t)=A_{\pm}\left(t-t_*\right)^2+\frac{N^2}{A_{\pm}},
\quad
\dot p_{\rm cl}(t)=2A_{\pm}\left(t-t_*\right),
\label{eq:pcl}
\end{equation}
where $A_{\pm}=\alpha/4$ and $t_*$ are fixed by the endpoints,
\begin{equation}
A_{\pm}=\frac{p_0+p_1\pm 2\sqrt{p_0p_1-N^2T^2}}{T^2},
\quad
t_* - t_0=\frac{T}{2}-\frac{p_1-p_0}{2A_{\pm}T}.
\label{eq:A-tstar}
\end{equation}
Accordingly, we obtain 
\begin{align}
\dot p_{\rm cl}(t_0)=-A_{\pm}T+\frac{p_1-p_0}{T},\quad 
\dot p_{\rm cl}(t_1)=A_{\pm}T+\frac{p_1-p_0}{T}.
\end{align}
Since the functional constraint enforces $p(t)=p_{\rm cl}(t)$, the on-shell action can be wirtten as,
\begin{align}
\label{eq:S-onshell}
\begin{split}
&S_\textrm{on-shell}[N]
= 2\pi^2N\!\int_{t_0}^{t_1}\mathrm{d}t\left(4-\Lambda p_{\rm cl}(t)\right)
+\frac{\pi^2}{N}\Bigl[q_0\dot p_{\rm cl}(t_0)-q_1\dot p_{\rm cl}(t_1)\Bigr] \\
&=2\pi^2N\left[4T-\Lambda\left(\frac{A_{\pm}T^3}{12}
+\frac{(p_1-p_0)^2}{4A_{\pm}T}+\frac{N^2T}{A_{\pm}}\right)\right] 
+\frac{\pi^2}{N}\left[\frac{p_1-p_0}{T}(q_0-q_1)-A_{\pm}T(q_0+q_1)\right].
\end{split}
\end{align}
Thus, we get 
\begin{align}
\begin{split}\label{eq:S-onshell-plus}
S_{\textrm{on-shell}}^{(+)}[N]
&=2\pi^2N\left[4T-\frac{\Lambda T}{3}\left(p_0+p_1-\sqrt{p_0p_1-N^2T^2}\right)\right] \\
&-\frac{2\pi^2}{NT}\left[(q_0+q_1)\sqrt{p_0p_1-N^2T^2}+p_0q_0+p_1q_1\right]\,.    
\end{split}
\end{align}
On the other hand, when we choose the $-$ sign, the on-shell action is given by 
\begin{align}
\begin{split}\label{eq:S-onshell-minus}
S_{\textrm{on-shell}}^{(-)}[N]
&=2\pi^2N\left[4T-\frac{\Lambda T}{3}\left(p_0+p_1+\sqrt{p_0p_1-N^2T^2}\right)\right] \\
&+\frac{2\pi^2}{NT}\left[(q_0+q_1)\sqrt{p_0p_1-N^2T^2}
-p_0q_0-p_1q_1\right]\,.    
\end{split}
\end{align}
which coincides with the on-shell action obtained in
Ref.~\cite{DiazDorronsoro:2018wro}.

\subsection{Functional Determinant}

We now expand $p=p_{\rm cl}+\delta p$ and linearize the constraint $\mathcal{G}[p]$ around the classical solution $p_{\rm cl}$.
Recalling the constraint 
\begin{equation}
\mathcal{G}[p]=-\frac{1}{4N}\Bigl(\frac{\dot p^{\,2}}{p}-2\ddot p\Bigr)-\frac{N}{p}\,,
\end{equation}
its first variation at $p_{\rm cl}$ is obtained by using
\begin{equation}
\delta\!\left(\frac{\dot p^{\,2}}{p}\right)
=\frac{2\dot p_{\rm cl}}{p_{\rm cl}}\,\delta\dot p
-\frac{\dot p_{\rm cl}^{2}}{p_{\rm cl}^{2}}\,\delta p,
\quad
\delta\!\left(\ddot p\right)=\delta\ddot p,
\quad
\delta\!\left(-\frac{N}{p}\right)=\frac{N}{p_{\rm cl}^2}\,\delta p,
\end{equation}
where we also used $(p_{\rm cl}+\delta p)^{-1}\simeq p_{\rm cl}^{-1}\left(1-\delta p/p_{\rm cl}\right)$.
Then, we get
\begin{align}
\begin{split}
\delta\mathcal{G}
&=-\frac{1}{4N}\left(
2\frac{\dot p_{\rm cl}}{p_{\rm cl}}\delta\dot p
-\frac{\dot p_{\rm cl}^{2}}{p_{\rm cl}^{2}}\delta p
-2\delta\ddot p\right)
+\frac{N}{p_{\rm cl}^{2}}\delta p
\\
&=\left[
\frac{1}{2N}\partial_t^2
-\frac{1}{2N}\frac{\dot p_{\rm cl}}{p_{\rm cl}}\partial_t
+\left(\frac{1}{4N}\frac{\dot p_{\rm cl}^{2}}{p_{\rm cl}^{2}}+\frac{N}{p_{\rm cl}^{2}}\right)
\right]\delta p
\;\equiv\;\mathcal{O}\,\delta p.
\end{split}
\label{eq:Bianchi-determinant}
\end{align}

Next, we consider the corresponding Gelfand-Yaglom problem.
Let $y(t)$ solve $\mathcal{O}y=0$ with the boundary conditions.
Multiplying $\mathcal{O}y=0$ by $2N$ gives
\begin{equation}
\ddot{y}-\frac{\dot p_{\rm cl}}{p_{\rm cl}}\dot{y}
+\left(\frac{1}{2}\frac{\dot p_{\rm cl}^{2}}{p_{\rm cl}^{2}}+\frac{2N^2}{p_{\rm cl}^{2}}\right)y=0,
\quad
y(t_0)=0,\quad \dot{y}(t_0)=1.
\label{eq:GY-eq}
\end{equation}
To apply the standard Gelfand-Yaglom theorem, it is convenient to remove the first-derivative term
and rewrite the operator in the Schr\"odinger form.
We shall perform the field redefinition
\begin{equation}
y(t)=\sqrt{p_{\rm cl}(t)}\,\psi(t).
\label{eq:sch-transform}
\end{equation}
Substituting \eqref{eq:sch-transform} into \eqref{eq:GY-eq}, 
we find that $\psi(t)$ satisfies
\begin{equation}
\ddot\psi(t)+W(t)\,\psi(t)=0,
\quad
W(t)=\frac{\ddot p_{\rm cl}}{2p_{\rm cl}}
-\frac{\dot p_{\rm cl}^{\,2}}{4p_{\rm cl}^{\,2}}
+\frac{2N^2}{p_{\rm cl}^{\,2}}.
\label{eq:psi-eq}
\end{equation}

The boundary conditions also transform. From $y(t_0)=0$ we obtain $\psi(t_0)=0$.
Moreover, since $\psi(t_0)=0$,
\begin{equation}
\dot y(t_0)=\frac{p_{\rm cl}(t_0)}{2\sqrt{p_{\rm cl}(t_0)}}\psi(t_0)+
\sqrt{p_{\rm cl}(t_0)}\,\dot\psi(t_0)=
\sqrt{p_0}\,\dot\psi(t_0),
\end{equation}
where $p_0:=p_{\rm cl}(t_0)$.
For the standard Gelfand-Yaglom normalization, we therefore define
\begin{equation}
Y(t):=\sqrt{p_0}\,\psi(t),
\quad
Y(t_0)=0,\quad \dot Y(t_0)=1\,.
\label{eq:GY-standard-IC}
\end{equation}
A first nontrivial solution of Eq.~\eqref{eq:psi-eq} 
is obtained from $y_1(t)=\dot p_{\rm cl}(t)$.
Indeed, by using Eq.~\eqref{eq:sch-transform}, it corresponds to
\begin{equation}
\psi_1(t)=\frac{y_1(t)}{\sqrt{p_{\rm cl}(t)}}=\frac{\dot p_{\rm cl}(t)}{\sqrt{p_{\rm cl}(t)}}.
\end{equation}
Since Eq.~\eqref{eq:psi-eq} has no first-derivative term, we obtain the following independent solution as
\begin{equation}
\psi_2(t)=C\,\psi_1(t)\int^{t}\!ds\,\frac{1}{\psi_1(s)^2},
\end{equation}
with an integration constant $C$.
Choosing the lower limit at $t_0$ and imposing $Y(t_0)=0$ yields
\begin{equation}
Y(t)
=\frac{\dot p_{\rm cl}(t_0)}{\sqrt{p_0}}\,
\frac{\dot p_{\rm cl}(t)}{\sqrt{p_{\rm cl}(t)}}
\int_{t_0}^{t}\!ds\,
\frac{p_{\rm cl}(s)}{\dot p_{\rm cl}(s)^2},
\label{eq:PsiGY-def}
\end{equation}
which indeed satisfies $\dot Y(t_0)=1$.
Therefore, we obtain
\begin{equation}
Y(t_1)
=\frac{\dot p_{\rm cl}(t_0)\dot p_{\rm cl}(t_1)}{\sqrt{p_0p_1}}
\int_{t_0}^{t_1}\!ds\,
\frac{p_{\rm cl}(s)}{\dot p_{\rm cl}(s)^2}\,.
\label{eq:PsiGY-1}
\end{equation}
where $p_1:=p_{\rm cl}(t_1)$.
Using the classical solution~\eqref{eq:pcl} and defining $s:=t-t_*$,
\begin{equation}
s_0:=t_0-t_*=-\frac{T}{2}+\frac{p_1-p_0}{2A_{\pm}T},
\quad
s_1:=t_1-t_*=\frac{T}{2}+\frac{p_1-p_0}{2A_{\pm}T},
\end{equation}
we obtain
\begin{equation}
\int_{s_0}^{s_1}\!ds\,\frac{p_{\rm cl}(s)}{\dot p_{\rm cl}(s)^2}
=\frac{T}{4A_{\pm}}+\frac{N^2}{4A_{\pm}^3}
\left(\frac{1}{s_0}-\frac{1}{s_1}\right)\,.
\label{eq:int-exact}
\end{equation}
Using $\dot p_{\rm cl}(t_0)=2A_{\pm}s_0$ and $\dot p_{\rm cl}(t_1)=2A_{\pm}s_1$ in Eq.~\eqref{eq:PsiGY-1}, we obtain
\begin{align}
\begin{split}
Y(t_1)
&=\frac{(2A_{\pm}s_0)(2A_{\pm}s_1)}{\sqrt{p_0p_1}}
\left[
\frac{T}{4A_{\pm}}+\frac{N^2}{4A_{\pm}^3}\left(\frac{1}{s_0}-\frac{1}{s_1}\right)
\right] \\
&=\frac{1}{\sqrt{p_0p_1}}
\left[
\frac{(p_1-p_0)^2}{4A_{\pm}T}
-\frac{A_{\pm}T^3}{4}
+\frac{N^2T}{A_{\pm}}
\right] \\
&=\mp\,\frac{T}{\sqrt{p_0p_1}}\sqrt{p_0p_1-N^2T^2}\,,
\end{split}
\label{eq:PsiGY-final}
\end{align}
where we used $s_0s_1=\frac{(p_1-p_0)^2}{4A_{\pm}^2T^2}-\frac{T^2}{4}$ and $s_1-s_0=T$.

Up to an overall normalization constant $\mathcal{C}_\pm$,
the determinant contributes $1/Y(t_1)$ to the prefactor.
Therefore, the propagator takes the form
\begin{equation}
\label{eq:Bianchi-propagator}
\begin{split}
K(q_1,p_1;q_0,p_0)
&=\sum_{\pm}\;
\frac{\mathcal{C}_{\pm}}{Y(t_1)}\,
\exp\!\left[\frac{i}{\hbar}S_{\textrm{on-shell}}^{(\pm)}[N]\right] \\
&=\sum_{\pm}\;
\frac{\mp\sqrt{p_0p_1}\,\mathcal{C}_{\pm}}{T\sqrt{p_0p_1-N^2T^2}}\,
\exp\!\left[\frac{i}{\hbar}S_{\textrm{on-shell}}^{(\pm)}[N]\right],    
\end{split}
\end{equation}
where $S_{\textrm{on-shell}}^{(\pm)}[N]$ is given in Eqs.~\eqref{eq:S-onshell-plus} and \eqref{eq:S-onshell-minus}. In the same way as \ac{jt} gravity case, the normalization constant $\mathcal{C}_{\pm}$ can be fixed by the composition law
\begin{align} 
\int \mathrm{d}q\, \mathrm{d}p\; K(q_1,p_1;q,p)\;
K(q,p;q_0,p_0)
= K(q_1,p_1;q_0,p_0) \,. 
\end{align}
By using the composition law, we obtain 
\begin{equation}
\mathcal{C}_\pm
=-\frac{\pi}{2\hbar N}\,.
\end{equation}
The fixed-lapse propagator can be expressed as
\begin{equation}\label{eq:Bianchi-propagator-final}
K(q_1,p_1;q_0,p_0)=\sum_{\pm}\;
\frac{\pi}{2\hbar NT}\frac{\sqrt{p_0p_1}}{\sqrt{p_0p_1-N^2T^2}}\,
\exp\!\left[\frac{i}{\hbar}S_{\textrm{on-shell}}^{(\pm)}[N]\right]\,,
\end{equation}
which agrees with the propagator obtained in
Ref.~\cite{DiazDorronsoro:2018wro}.
\footnote{
Our convention keeps the overall factor $2\pi^2$ in the action~\eqref{eq:BIX-action}, whereas Ref.~\cite{DiazDorronsoro:2018wro} absorbs this factor into its normalization conventions.  Therefore, by shifting $\hbar \to 2\pi^2\hbar$, we obtain an expression that matches exactly.
}
\footnote{
According to the author's comment, the prefactor in Ref.~\cite{DiazDorronsoro:2018wro} was given by the Van Vleck determinant formula.
}

In summary, evaluating the functional determinant via the Gelfand-Yaglom theorem fixes the prefactor and thereby validates our treatment of the constrained path integral.
Finally, integrating the fixed-lapse propagator~\eqref{eq:Bianchi-propagator-final} over the positive lapse yields the full Lorentzian transition amplitude.
A detailed analysis of the associated thimble structure for the Bianchi~IX lapse integral, including several different boundary conditions, can be found in
Refs.~\cite{DiazDorronsoro:2018wro,Feldbrugge:2018gin,Janssen:2019sex,Lehners:2024kus}, so we will not discuss it further here.

\section{Conclusion}
\label{sec:conclusion}

In this work, we evaluated minisuperspace path integrals in \ac{jt} gravity and Bianchi~IX quantum cosmology using functional determinant methods. In both models, integrating out an auxiliary minisuperspace variable transforms the original path integral into a constrained one: a functional delta function enforces the classical constraint equation, thereby localizing the remaining integral to classical configurations. Quantum fluctuations enter through the corresponding Jacobian, namely the functional determinant of the operator obtained by linearizing the constraint. We computed this determinant exactly using the Gelfand-Yaglom theorem and fixed the overall normalization by imposing the composition law, thereby obtaining properly normalized fixed-lapse propagators.

For \ac{jt} gravity, the constrained formulation reproduces the propagator previously obtained by performing the dilaton and scale-factor path integrals explicitly~\cite{Honda:2024hdr}, providing a nontrivial consistency check of the delta-functional treatment and the associated prefactor. Upon integrating over the lapse, we derived closed-form expressions in terms of Bessel/Hankel functions and reviewed the resulting saddle-point structure~\cite{Honda:2024hdr}, which separates an oscillatory, semiclassical regime from an exponentially suppressed, genuinely quantum regime. We also extended the analysis to a quadratic dilaton potential $U(\phi)=m^{2}\phi^{2}$. In this case, the determinant prefactor is unchanged, and the qualitative thimble structure is preserved.

As an application beyond \ac{jt} gravity, we applied the same strategy to biaxial Bianchi~IX minisuperspace. Integrating out one minisuperspace variable produces a functional constraint for the remaining scale factor degree of freedom. We evaluated the associated functional determinant via the Gelfand-Yaglom method and obtained the fixed-lapse propagator including its prefactor. Our result agrees with the propagator reported in Ref.~\cite{DiazDorronsoro:2018wro}.

Our results support the reliability of the constrained path integral approach: once the functional determinant is treated exactly, the method yields correctly normalized transition amplitudes and reproduces known results. Finally, while the constrained expression \eqref{eq:jt-amplitude} introduced for \ac{jt} gravity is well-defined in minisuperspace, extending the determinant computation to fluctuations of the full spacetime metric $g_{\mu\nu}$ beyond minisuperspace requires a substantially more complicated analysis. We leave this fully covariant extension to future work.

\section*{Acknowledgments}
H. M. would like to thank Masazumi Honda, Oliver Janssen, Kota Numajiri,  Kazumasa Okabayashi, Daiki Saito, and Takahiro Terada for useful discussions regarding the minisuperspace analysis of \ac{jt} gravity and Bianchi IX quantum cosmology. This work is supported by JSPS KAKENHI Grant No. JP23K13100.

\bibliography{Refs}
\bibliographystyle{JHEP}

\end{document}